\documentclass[twocolumn,showpacs,preprintnumbers,amsmath,amssymb]{revtex4}
\usepackage{graphicx}
\usepackage{dcolumn}
\usepackage{bm}
\begin{document}
\title{Direct measurements of the fractional quantum Hall effect gaps}
\author{V.~S. Khrapai, A.~A. Shashkin, M.~G. Trokina, and V.~T. Dolgopolov}
\affiliation{Institute of Solid State Physics, Chernogolovka, Moscow
District 142432, Russia}
\author{V. Pellegrini and F. Beltram}
\affiliation{NEST INFM-CNR, Scuola Normale Superiore, Piazza dei
Cavalieri 7, I-56126 Pisa, Italy}
\author{G. Biasiol and L. Sorba$^*$}
\affiliation{NEST INFM-CNR and Laboratorio Nazionale TASC-INFM,
I-34012 Trieste, Italy\\
$^*$Scuola Normale Superiore, Piazza dei Cavalieri 7, I-56126 Pisa,
Italy}
\begin{abstract}
We measure the chemical potential jump across the fractional gap in
the low-temperature limit in the two-dimensional electron system of
GaAs/AlGaAs single heterojunctions. In the fully spin-polarized
regime, the gap for filling factor $\nu=1/3$ increases {\it linearly}
with magnetic field and is coincident with that for $\nu=2/3$,
reflecting the electron-hole symmetry in the spin-split Landau level.
In low magnetic fields, at the ground-state spin transition for
$\nu=2/3$, a correlated behavior of the $\nu=1/3$ and $\nu=2/3$ gaps
is observed.
\end{abstract}
\maketitle

Plateaux of the Hall resistance corresponding to zeros in the
longitudinal resistance at fractional filling factors of Landau
levels in two-dimensional (2D) electron systems, known as the
fractional quantum Hall effect \cite{tsui82}, are believed to be
caused by electron-electron interactions (for a recent review, see
Ref.~\cite{chakraborty00}). Two theoretical approaches to the
phenomenon have been formulated over the years. One approach is based
on a trial wave function of the ground state \cite{laughlin83}, and
the other exploits an introduction of composite fermions to reduce
the problem to a single-particle one \cite{jain89,halperin93}. The
fractional gap is predicted to be determined by the Coulomb
interaction in the form $e^2/\varepsilon l_B$ (where $\varepsilon$ is
the dielectric constant and $l_B=(\hbar c/eB)^{1/2}$ is the magnetic
length), which leads to a square-root dependence of the gap on
magnetic field, $B$. Observation of such a behavior would confirm the
predicted gap origin.

Attempts to experimentally estimate the fractional gap value yielded
similar results, at least, in high magnetic fields
\cite{boebinger85,haug06,eisenstein92,hirji03,hirji05}. Still, the
expected dependence of the gap on magnetic field has not been either
confirmed or rejected. The problems with experimental verification
are as follows. Standard measurements of activation energy at the
longitudinal resistance minima allow one to determine the mobility
gap \cite{boebinger85,haug06} which may be different from the gap in
the spectrum. The data for the gap obtained by thermodynamic
measurements depended strongly on temperature \cite{eisenstein92};
for this reason, the magnetic field dependence of the gap may be
distorted.

In this paper, we report measurements of the chemical potential jump
across the fractional gap at filling factor $\nu=1/3$ and $\nu=2/3$
in the 2D electron system in GaAs/AlGaAs single heterojunctions using
a magnetocapacitance technique. We find that the gap, $\Delta\mu_e$,
increases with decreasing temperature and in the low-temperature
limit, it saturates and becomes independent of temperature. In
magnetic fields above $\approx5$~T, the limiting value,
$\Delta\mu_e^0$, for $\nu=1/3$ is described with good accuracy by a
linear increase of the gap with magnetic field and is practically
coincident with the gap for $\nu=2/3$. In lower magnetic fields, the
minimum of the gap $\Delta\mu_e^0$ at $\nu=2/3$ occurring at a
critical field of $\approx4$~T, which corresponds to ground-state
spin transition \cite{eisenstein90,clark90,engel92,qq97}, is
accompanied with a change in the behavior of that at $\nu=1/3$. The
correlation between the magnetic field dependences of both gaps
indicates the presence of a spin transition for the $\nu=1/3$ gap.
The linear dependence of the fractional gap on magnetic field as well
as the electron-hole symmetry in the spin-split Landau level are the
case in the completely spin-polarized regime. Since the
interaction-enhanced gaps in the integer quantum Hall effect in
different 2D electron systems also increase linearly with magnetic
field \cite{dolgopolov97,khrapai03}, the linear law seems robust.
This indicates that electron-electron interactions in 2D should be
treated in a less straightforward way.

Measurements were made in an Oxford dilution refrigerator with a base
temperature of $\approx30$~mK on remotely doped GaAs/AlGaAs single
heterojunctions with a low-temperature mobility $\approx4\times
10^6$~cm$^2$/Vs at electron density $9\times 10^{10}$~cm$^{-2}$.
Samples had the quasi-Corbino geometry with areas $27\times 10^4$
(sample 1) and $2.1\times 10^4$~$\mu$m$^2$ (sample 2). A metallic
gate was deposited onto the surface of the sample, which allowed
variation of the electron density by applying a dc bias between the
gate and the 2D electrons. The gate voltage was modulated with a
small ac voltage of 2.5~mV at frequencies in the range 0.1--2.5~Hz,
and both the imaginary and real components of the current were
measured with high precision ($\sim10^{-16}$~A) using a
current-voltage converter and a lock-in amplifier. Smallness of the
real current component as well as proportionality of the imaginary
current component to the excitation frequency ensure that we reach
the low-frequency limit and the measured magnetocapacitance is not
distorted by lateral transport effects. A dip in the
magnetocapacitance in the quantum Hall effect is directly related to
a jump of the chemical potential across a corresponding gap in the
spectrum of the 2D electron system \cite{smith85}:
\begin{equation}\frac{1}{C}=\frac{1}{C_0}+\frac{1}{Ae^2dn_s/d\mu},\label{C}\end{equation}
where $C_0$ is the geometric capacitance between the gate and the 2D
electrons, $A$ is the sample area, and the derivative $dn_s/d\mu$ of
the electron density over the chemical potential is the thermodynamic
density of states.

\begin{figure}
\scalebox{0.38}{\includegraphics[clip]{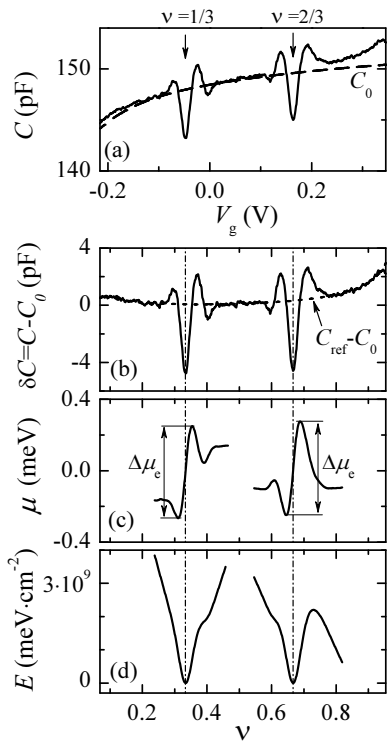}}
\caption{\label{fig1} (a)~Magnetocapacitance as a function of gate
voltage in sample 1 in a magnetic field of 9~T at a temperature of
0.18~K. Also shown by a dashed line is the geometric capacitance
$C_0$. (b)~The difference $C-C_0$ as a function of filling factor.
The dotted line corresponds to the reference curve $C_{\text{ref}}$.
(c)~The chemical potential near $\nu=1/3$ and $\nu=2/3$ obtained by
integrating the magnetocapacitance; see text. (d)~The energy of the
2D electron system near $\nu=1/3$ and $\nu=2/3$ obtained by
integrating the chemical potential. The zero level in (c) and (d)
corresponds to the fractional $\nu$.}
\end{figure}

A magnetocapacitance trace $C$ as a function of gate voltage $V_g$ is
displayed in Fig.~\ref{fig1}(a) for a magnetic field of 9~T. Narrow
minima in $C$ accompanied at their edges by local maxima are seen at
filling factor $\nu\equiv n_shc/eB=1/3$ and $\nu=2/3$. Near the
filling factor $\nu=1/2$, the capacitance $C$ in the range of
magnetic fields studied reaches its high-field value determined by
the geometric capacitance $C_0$ (dashed line). We have verified that
the obtained $C_0$ corresponds to the value calculated using
Eq.~(\ref{C}) from the zero-field capacitance and the density of
states $m/\pi\hbar^2$ (where $m=0.067m_e$ and $m_e$ is the free
electron mass). The geometric capacitance $C_0$ increases with
electron density as the 2D electrons are forced closer to the
interface. The chemical potential jump $\Delta\mu_e$ for electrons at
fractional filling factor can be determined by integrating the
magnetocapacitance over the dip:
\begin{equation}\Delta\mu_e=\frac{Ae^2}{C_0}\int_{\text{dip}}\frac{C_0-C}{C}dn_s=\frac{e}{C_0}\int_{\text{dip}}(C_0-C)dV_g.\label{Delta}\end{equation}
As seen from Fig.~\ref{fig1}(b), the difference $\delta C=C-C_0$
versus $\nu$ is nearly symmetric about filling factor $\nu=1/2$.
Maxima in $\delta C>0$ observed near $\nu=0$ and $\nu=1$ are related
to the so-called negative thermodynamic compressibility
\cite{efros88,macdonald86}. The effect is caused by the intra-level
interactions between quasiparticles which provide a negative
contribution of order $-(e^2/\varepsilon l_B)\{\nu\}^{1/2}$ to the
chemical potential (here $\{\nu\}$ is the deviation of the filling
factor from the nearest integer). This leads to an upward shift of
the features at fractional filling factors which is more pronounced
for $\nu=2/3$ and for magnetic fields below $\approx10$~T. To allow
for the shift, we replace the value $C_0$ in the integrand of
Eq.~(\ref{Delta}) by a reference curve $C_{\text{ref}}$ that is
obtained by extrapolating the $C(\nu)$ dependence from above and
below the fractional-$\nu$ feature. Based on the above contribution
$d\mu/dn_s\propto\{\nu\}^{-1/2}$ in the spirit of
Ref.~\cite{eisenstein92}, one can describe the experimental $\delta
C(\nu)$ excluding the fractional-$\nu$ structures (the dotted line in
Fig.~\ref{fig1}(b)). This naturally gives an extrapolation curve
$C_{\text{ref}}$. Importantly, we have verified that the determined
$\Delta\mu_e$ is not sensitive to the particular extrapolation law.
Note that unlike penetrating field technique \cite{eisenstein92}, our
method is not able to yield the thermodynamic compressibility in
absolute units; however, both techniques are equally good for
determining gaps in the spectrum.

\begin{figure}
\scalebox{0.37}{\includegraphics[clip]{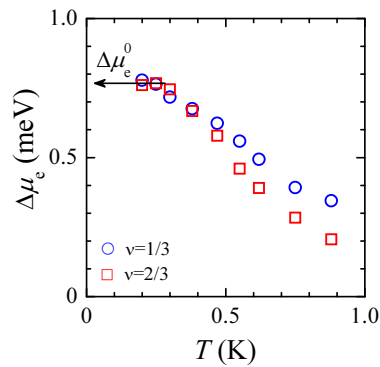}}
\caption{\label{fig2} Dependence of the fractional gap on temperature
in sample 2 at $B=11.5$~T. The value $\Delta\mu_e^0$ is indicated.}
\end{figure}

It is easy to determine the behavior of the chemical potential when
the filling factor traverses the fractional gap, as shown in
Fig.~\ref{fig1}(c). The chemical potential jump corresponds to a cusp
on the dependence of the energy, $E$, of the 2D electron system on
filling factor \cite{chakraborty00}, shown in Fig.~\ref{fig1}(d) for
illustration. The difference between the $\mu$ values well above and
well below the jump is smaller than $\Delta\mu_e$, as determined by
the local maxima at the edges of the dip in $C$. This can be caused
by both the gap closing and the interaction effect similar to that
mentioned above for integer filling factor. The magnetocapacitance
data do not allow one to distinguish whether or not the fractional
gap closes when the Fermi level lies outside the gap.

In Fig.~\ref{fig2}, we show the temperature dependence of the gap for
$\nu=1/3$ and $\nu=2/3$. As the temperature is decreased, the value
$\Delta\mu_e$ increases and in the limit of low temperatures, the gap
saturates and becomes independent of temperature. It is the saturated
low-temperature value $\Delta\mu_e^0$ that will be studied in the
following as a function of the magnetic field.

\begin{figure}
\scalebox{0.39}{\includegraphics[clip]{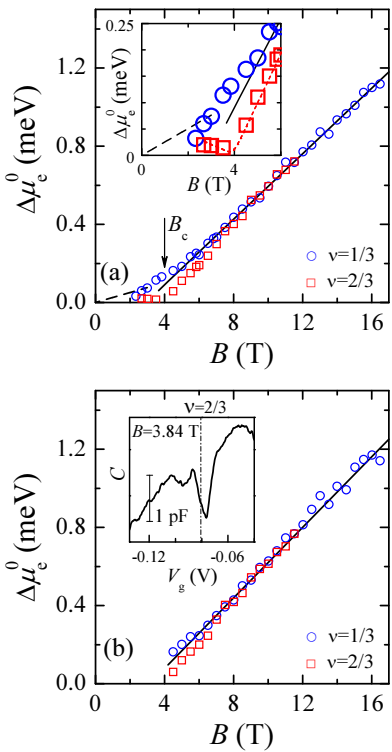}}
\caption{\label{fig3} (a)~Change of the fractional gap
$\Delta\mu_e^0$ with magnetic field for sample 1. The solid line is a
linear fit to the high-field data for the $\nu=1/3$ gap. The change
in the behavior of the $\nu=1/3$ gap is marked by the arrow. The
dashed line corresponds to the Zeeman energy in bulk GaAs. A close-up
view on the low-field region is displayed in the inset. The dotted
line is a guide to the eye. (b)~The same as in (a) for sample 2.
Inset: magnetocapacitance versus gate voltage in sample 1 at
$T\approx30$~mK.}
\end{figure}

In Fig.~\ref{fig3}, we show how the $\nu=1/3$ and $\nu=2/3$ gap
$\Delta\mu_e^0$ changes with magnetic field. Above $B\approx5$~T, the
data for $\nu=1/3$ are best described by a linear increase of the gap
with magnetic field \cite{remark} and are practically coincident with
the data for $\nu=2/3$. In lower magnetic fields, the value
$\Delta\mu_e^0$ for $\nu=2/3$ is a minimum (inset to
Fig.~\ref{fig3}(a)), which corresponds to spin transition in the
ground state \cite{eisenstein90,clark90,engel92,qq97}. The occurrence
of the spin transition has been double checked by analyzing the
$C(V_g)$ curves (inset to Fig.~\ref{fig3}(b)). Just below $B=4$~T, in
the close vicinity of the transition, a double-minimum structure in
the capacitance at filling factor $\nu=2/3$ is observed. This
structure is conspicuous at the lowest temperatures, as caused by the
resistive contribution to the minima in $C$, and is similar to the
structures observed at the magnetic transitions in the integer
\cite{khrapai00,poortere00,muraki01} and $\nu=2/3$ fractional quantum
Hall effect \cite{clark90,engel92,dolgopolov06}. As seen from
Fig.~\ref{fig3}(a), the minimum of the $\nu=2/3$ gap is accompanied
with a deviation at $B=B_c\approx4$~T in the magnetic field
dependence of the $\nu=1/3$ gap from the linear behavior, the
lowest-$B$ data for the latter gap being comparable to the Zeeman
energy in bulk GaAs (dashed line). It is worth noting that the data
obtained on both samples in the magnetic field range where they
overlap are very similar; measurements on sample 2 in magnetic fields
below 4~T were hampered by the integer quantum Hall effect in
Corbino-like contacts.

Since in high magnetic fields, the value of fractional gap determined
in the experiment is similar to the previously obtained results
\cite{boebinger85,haug06,eisenstein92}, it is unlikely to be strongly
influenced by the residual disorder in the 2D electron system.
Bearing in mind the above mentioned problems with the data
interpretation of transport experiments, one can compare the gap
$\Delta\mu_e^0$ obtained by thermodynamic measurements with the
recent results of activation energy measurements in similar samples
at $\nu=1/3$ \cite{haug06}. The double activation energy,
$\Delta_{\text{act}}$, is approximately equal to $\Delta\mu_e^0$ at
$B\lesssim8$~T, whereas in higher magnetic fields, its increase with
$B$ weakens compared to that of $\Delta\mu_e^0(B)$. Although the
ratio $\Delta\mu_e^0/\Delta_{\text{act}}$ reaches the factor of about
1.7 in the highest accessible magnetic fields, it is still far from
the theoretically expected value
$\Delta\mu_e^0/\Delta_{\text{act}}=3$ \cite{halperin93,gros90}. So,
the obtained data do not confirm the prediction that the excitations
at filling factor $\nu=1/3$, which are relevant for activation energy
measurements, are quasiparticles with fractional charge $e/3$.

We would like to emphasize that the thermodynamic measurements allow
us to study the variation of the ground-state energy of the 2D
electron system as the density of quasiparticles is varied at filling
factors around $\nu=1/3$ and $\nu=2/3$; the fractional gap is equal
to
$\Delta\mu_e(\nu)=\left.dE/dn_s\right|_{\nu+0}-\left.dE/dn_s\right|_{\nu-0}$.
As inferred from the correlated behavior of the gaps at $\nu=1/3$ and
$\nu=2/3$ in low magnetic fields (Fig.~\ref{fig3}(a)), the change at
$B=B_c$ in the dependence of the $\nu=1/3$ gap on magnetic field is
connected with the expected change of the spin of the ``excited''
state \cite{chakraborty00,mariani02}, i.e., the 2D electron system at
filling factors just above $\nu=1/3$ should be fully spin-polarized
at $B>B_c$, while in the opposite case of $B<B_c$, the $\nu=1/3$ gap
is expected to be of Zeeman origin. The regime of small Zeeman
energies was studied in Ref.~\cite{leadley97} where the
pressure-induced vanishing of the $g$-factor/Zeeman energy in
$B\gtrsim5$~T was found to be responsible for the decrease of the
$\nu=1/3$ gap. Note that the lowest-$B$ behavior may be more
sophisticated because the gap can in principle collapse due to
disorder in the 2D electron system. In the high-field limit, we are
interested in here, the gaps at $\nu=1/3$ and $\nu=2/3$ are
coincident with each other reflecting the electron-hole symmetry in
the spin-split Landau level (Figs.~\ref{fig1} and \ref{fig3}), and
the electron spins are completely polarized both at $\nu$ just above
1/3 and at $\nu$ just above 2/3.

It is important that the linear dependence of the fractional gap on
magnetic field is observed in the fully spin-polarized regime. This
linear behavior is concurrent with that of the interaction-enhanced
gaps in the integer quantum Hall effect \cite{dolgopolov97,khrapai03}
and, therefore, the linear law seems robust being valid for different
gaps of many-body origin in different 2D electron systems with
different interaction and disorder strengths. This points to a
failure of the straightforward way of treating electron-electron
interactions in 2D which leads to a square-root dependence of the gap
on magnetic field. There is a noteworthy distinction between the
interaction-enhanced gaps in the integer quantum Hall effect and the
fractional gap: while in the former case the exchange energy is
involved due to a change of the (iso)spin index, the exchange energy
contribution to the fractional gap in the fully spin-polarized regime
is not expected. Most likely, electron-electron correlations in 2D
systems should be considered more carefully for the sufficient theory
to be developed.

In summary, we have studied the variation of the ground-state energy
of the 2D electron system in GaAs/AlGaAs single heterojunctions with
changing filling factor around $\nu=1/3$ and $\nu=2/3$ and determined
the fractional gap in the limit of low temperatures. In low magnetic
fields, the minimum of the $\nu=2/3$ gap, which corresponds to
ground-state spin transition, is found to be accompanied with a
change in the behavior of the $\nu=1/3$ gap. The correlation between
the dependences of both gaps on magnetic field indicates the presence
of a spin transition for the $\nu=1/3$ gap. In high magnetic fields,
in the completely spin-polarized regime, the gap for $\nu=1/3$
increases linearly with magnetic field and is coincident with that
for $\nu=2/3$, reflecting the electron-hole symmetry in the
spin-split Landau level. The straightforward way of treating
electron-electron interactions in 2D fails to explain the linear
behavior of the fractional gap which is concurrent with that of the
interaction-enhanced gaps in the integer quantum Hall effect.

We gratefully acknowledge discussions with S.~V. Kravchenko. We would
like to thank J.~P. Kotthaus for an opportunity to use the clean room
facilities at LMU Munich and C. Dupraz for technical assistance. This
work was supported by the RFBR, RAS, and the Programme ``The State
Support of Leading Scientific Schools''. VSK acknowledges the A.~von
Humboldt foundation.

\end{document}